\documentclass[aps,pre,twocolumn,showpacs]{revtex4}
\usepackage{graphicx}
\begin{document}
\title{{\hfill \small \tt Phys. Rev. E {\bf 84}, 031102 (2011)} \\
Glassy phases and driven response of the phase-field-crystal
model with random pinning}

\author{E. Granato $^{1,2}$, J.A.P. Ramos $^{1,3}$, C.V. Achim $^{4,5}$, J. Lehikoinen $^{4}$, S.C. Ying $^2$,
T. Ala-Nissila $^{2,4}$, and K.R. Elder $^6$}

\address{$^1$Laborat\'orio Associado de Sensores e Materiais,
Instituto Nacional de Pesquisas Espaciais,12227-010 S\~ao Jos\'e dos
Campos, SP, Brazil}
\address{$^2$Department of Physics, P.O. Box 1843, Brown University,
Providence, RI 02912-1843, USA}
\address{$^3$ Departamento de Ci\^encias Exatas,
Universidade Estadual do Sudoeste da Bahia, 45000-000 Vit\'oria da
Conquista, BA,Brazil}
\address{$^4$Department of Applied Physics,
Aalto University School of Science, P.O. Box 11000,
FI-00076 Aalto, Espoo, Finland}
\address{$^5$
Institut f\"{u}r Theoretische Physik II: Weiche Materie,
Heinrich-Heine-Universit\"{a}t D\"{u}sseldorf,
Universit\"{a}tsstra{\ss}e 1, D-40225 D\"{u}sseldorf, Germany}
\address{$^6$Department of Physics, Oakland University, Rochester,
Michigan 48309-4487, USA}

%\date{November 9, 2010}

\begin{abstract}
We study %structural correlations the ground state properties
the structural correlations and the nonlinear response to a driving
force of a two-dimensional phase-field-crystal model with random
pinning. The model provides an effective continuous  description of
lattice systems in the presence of  disordered external pinning
centers, allowing for both elastic and plastic deformations. We find
that the phase-field crystal with disorder assumes an amorphous
glassy ground state, with only short-ranged positional and
orientational correlations even in the limit of weak disorder. Under
increasing driving force, the pinned amorphous-glass phase evolves
into a moving plastic-flow phase and then finally a moving smectic
phase. The transverse response of the moving smectic phase  shows a
vanishing transverse critical force for increasing system sizes.
\end{abstract}


\maketitle

\section{Introduction}

Pinning and sliding of lattice systems, which can form periodic
structures in the absence of perturbations, are subjects of
considerable interest. In the presence of pinning disorder and
driving forces, they can exhibit a wide variety of interesting
equilibrium and non-equilibrium behavior with partially ordered and
glassy structures. Important examples in two dimensions include
vortex lattices in superconducting films \cite{martin97}, atomic
layers adsorbed between sliding surfaces \cite{Book} or on
oscillating substrates \cite{Mistura,gy04}, and colloidal crystals
on a rough substrate \cite{Pert01,Pert08}. Although in three
dimensions  a topological ordered phase, the Bragg glass
\cite{giamarchi95}, with quasi-long range positional order is
possible in the weak disorder regime, the two-dimensional limit is
qualitatively different due to the proliferation of thermally and
disorder-induced dislocations even in the weak-disorder regime.
Analytical and numerical studies for the equilibrium behavior
\cite{giamarchi95,carraro,carpentier,fisher99,doussal00,chandra}
have shown that, in two dimensions, positional and orientational
order are both destroyed by weak pinning disorder and topological
defects at any nonzero temperature leading to a liquid-like phase in
the thermodynamic limit. In the absence of thermal fluctuations
(zero temperature) an amorphous glass is expected
\cite{carpentier,giamarchi}. For the system under a driving force
moving at high velocities, it has been shown analytically that some
components of the disorder remain static in the co-moving reference
frame, leading to a moving glass phase \cite{giamarchi}. In
three-dimensional systems, such state can show topological order as
a moving Bragg glass \cite{giamarchi,balents,pardo}. In two
dimensions, however, this phase corresponds to a moving smectic
glass, which retains quasi long-range order in the direction
transverse to the driving force but only exponential correlations in
the parallel direction
\cite{giamarchi,balents,Moon,olson98,olson00,fanghor,reichardt,lefebfre}.

In modeling such lattice systems, both for static and dynamic
properties, it is essential to include the periodicity of the
lattice and allow for topological defects (dislocations and
disclinations). These topological defects are not contained in pure
elastic models that  have completely different properties, specially
in two dimensions.  A phase-field-crystal (PFC) model was introduced
recently \cite{Elder02,Elder04,Elder07} that allows for both elastic
deformations and topological defects within an effective continuous
description of the lattice system while still retaining information
on short length scales. By extending the PFC model to take into
account the effect of an external periodic pinning potential
\cite{achim06}, a two-dimensional version of the model has been used
to describe commensurate-incommensurate transitions in the presence
of thermal fluctuations \cite{ramos08} and the driven response
\cite{achim09} including inertial effects \cite{Ramos10}. However,
in order to study the statics and dynamics of the disordered system,
quenched pinning disorder needs to be included in the PFC modeling.

In this work, we study %structural correlations the ground state properties
the structural correlations and nonlinear response to a driving
force of a two-dimensional PFC model with random pinning. The model
provides an effective continuous description of lattice systems in
presence of disordered external pinning centers, allowing for both
elastic and plastic deformations. We show that in the presence of
disorder, the phase-field crystal assume an amorphous glassy ground
state, with only short-ranged positional and orientational
correlations even in the limit of weak disorder. Under increasing
driving force, the pinned amorphous-glass phase evolves into a
moving plastic-flow phase and then finally a smectic phase. The
transverse response of the moving smectic phase shows a vanishing
transverse critical force for increasing system sizes.

\section{PFC model with random pinning potential}

The effective Hamiltonian  of the PFC model with an external pinning
potential \cite{achim06} can be written in a dimensionless form as
\begin{equation}
H_{\rm pfc} = \int d \vec x \{  \frac{1}{2} \psi[r+(1+\nabla^2)^2]
\psi + \frac{\psi^4}{4} + V(\vec x) \psi (\vec x) \}, \label{pfcp}
\end{equation}
where $\psi(\vec x)$ is a continuous field at position $\vec x $ in
two dimensions and $V(\vec x)$ is the pinning potential.  The field
$\psi(\vec x)$ is conserved and its average value, $\bar \psi$,
together with $ r$ are the relevant parameters in the model.

In the absence of the pinning potential, $V(\vec x)=0$, the  energy
functional of Eq. (\ref{pfcp}) can be minimized by a configuration
of the field $\psi(\vec x)$ forming a hexagonal pattern of peaks
with  wave vector of magnitude $|\vec Q| \approx 1$, when the values
of the parameters $r$ and $\bar \psi$  are chosen appropriately.
This structure of peaks can be regarded as the ground state of a
lattice system with perfect crystalline order, where the phase field
$\psi(\vec x)$ represents the deviation of the particle number
density $\rho(\vec x)$ from a reference value $\rho_0$, $\psi(\vec
x)= (\rho(\vec x) - \rho_0)/\rho_0$. The energy functional of Eq.
(\ref{pfcp}) can then be used to describe both elastic as well
plastic properties of the lattice system \cite{Elder02,Elder04},
including the effects of thermal fluctuations \cite{ramos08}. An
external driving force $\vec f$ in the lattice system, can be
represented by an additional contribution to the effective
Hamiltonian as
\begin{equation}
H_f = - \int d \vec x \rho(\vec x) \vec x \cdot \vec f ,
\label{energy}
\end{equation}
which will lead to nonequilibrium behavior. The effects of the
driving force have been recently investigated in some detail
\cite{achim09,Ramos10}, in absence of disorder. Inertial effects of
the lattice system can also be described within the PFC model, by
including an additional kinetic energy contribution to the effective
Hamiltonian as
\begin{equation}
H_{kin} =  \int d \vec x \frac{\vec g ^2 (\vec x)}{2 \rho(\vec x)} ,
\label{kenergy}
\end{equation}
where $\vec g(\vec x)$ is the momentum density field.

The dynamical equations describing the time evolution of the lattice
system in presence of thermal fluctuations and the external force
can be written as \cite{Ramos10}

\begin{eqnarray}\label{cpeqs}
\frac{\partial \psi}{\partial t}  =  -\nabla \cdot \vec g; & &\\
\nonumber \frac{\partial g_i}{\partial t}= -\nabla_i
\frac{\delta H_{\rm pfc}}{\delta \psi} &+& \psi f_i - \gamma g_i + \nu_i(\vec x,t); \\
\nonumber \langle \nu_i(\vec x,t) \nu_j(\vec x^\prime, t^\prime)
\rangle &=& 2 k_B T \gamma \delta (\vec x - \vec x^\prime) \delta (t
- t^\prime)\delta_{i,j},
\end{eqnarray}
where $\vec f=(f_x,f_y)$ is the spatially uniform external force and
$\vec \nu(\vec x,t)$ is a thermal noise satisfying the
fluctuation-dissipation relation corresponding to a temperature $T$
and damping coefficient $\gamma$.

In the present work, we only consider the limit of very large
$\gamma$, when inertial effects are negligible, leading to the
overdamped dynamical equations \nonumber
\begin{eqnarray}\label{coveq}
\frac{\partial \psi(\vec x, t)}{\partial t}  =  -\nabla \cdot \vec g; & &\\
\nonumber
 \gamma g_i= -\nabla_i
\frac{\delta H_{\rm pfc}}{\delta \psi} &  + & \psi (\vec x,t) f_i  + \nu_i(\vec x,t); \\
\nonumber \langle \nu_i(\vec x,t) \nu_j(\vec x^\prime, t^\prime)
\rangle &=& 2 k_B T \gamma \delta (\vec x - \vec x^\prime) \delta (t
- t^\prime)\delta_{i,j},
\end{eqnarray}
%%
%% KEN - shouldn't the first two terms be divided by eta and
%% the last by sqrt(eta)?
%% ENZO - I put eta back in eqs. (5) and (6) and set eta=1 in sec. IV.
%% We can get rid of  \eta by redefining g, eta, t and psi
%% but it is confusing to explain and not too important at this point.
%%

The above coupled equations for $\psi$ and $\vec g$ can also be
combined into a single equation for $\psi$, giving
\begin{eqnarray}\label{soveq}
\gamma \frac{\partial \psi}{\partial t}  = \nabla^2
\frac{\delta H_{\rm pfc}}{\delta \psi}&-& \vec f \cdot \nabla \psi + \zeta(\vec x,t); \\
\nonumber
 \langle \zeta(\vec x,t) \zeta(\vec x^\prime, t^\prime) \rangle &=&
2 k_B T \gamma \nabla^2 \delta (\vec x - \vec x^\prime) \delta (t -
t^\prime).
\end{eqnarray}

The generalization of the PFC model with an external periodic
pinning potential studied previously
\cite{achim06,ramos08,achim09,Ramos10} to the case of a quenched
random pinning potential considered here is straightforward. Such a
model is relevant for studying diverse systems such as, adsorbate
layers with quenched impurities or on substrates with disorder, and
vortex lattice in superconductors in the presence of pinning
centers.
%%To this end, in the present work we consider two models of
%%random pinning. In the {\it dense} model the quenched potential
% is put at every (discretized) lattice site
To this end we model the quenched potential in the simplest way by
defining at every spatial location  $V(\vec x ) = D \mu(\vec x)$,
where $\mu(\vec x)$ is a $\delta$-correlated distribution
\begin{equation} \label{disor}
\langle \mu(\vec x) \mu(\vec x') \rangle = \delta(\vec x - \vec x'),
\end{equation}
and $D$ is an amplitude characterizing the strength of the disorder.
Since, there is no characteristic length scale in such pinning
potential, it is particularly suitable for the investigation of
finite-size effects using small system sizes as done in this work.
In this {\it dense} pinning model,
% in the numerical implementation of the PFC model about ten lattice
%points are needed to resolve the density peaks,
the separation between pinning centers can be much smaller than the
distance between the density peaks in the phase-field crystal. This
corresponds to a physical system where the length scale of the
varying pinning potential is smaller than the average lattice
spacing of the system. Such a scenario could be realized {\it e.g.}
in the case of colloidal particles confined near a  rough glass
plate \cite{Pert01,Pert08}.

For the numerical calculations, the phase field $\psi(\vec x)$ is
defined on a  space grid $(i dx, j dy)$ with periodic boundary
conditions. The simulations in presence of thermal fluctuations and
the driving force were performed using Eq. (\ref{coveq}), using
Euler's method with the Laplacians and gradients evaluated by finite
differences. In the absence of external force and thermal
fluctuations, the equation of motion (\ref{soveq}) was used in most
simulations, and solved using a semi-implicit algorithm. The linear
part is treated implicitly, while the non-linear term is treated
explicitly \cite{semiimplc}. The field at time $t+dt$ is obtained
according to
\begin{equation}\label{eq:siupalg}
 \hat \psi(\vec k,t+dt)=\frac{\hat \psi(\vec k,t)+(-k^2)\hat{NT}(\vec k,t)}{1-dt(-k^2)\hat \omega(\vec k)},
\end{equation}
where, $\hat \psi(\vec k,t)$ is the Fourier transform of $\psi(\vec
r,t)$, $-k^2$ is the equivalent in inverse space of the laplacian
$\nabla^2$ and $\omega(\vec k)=r+(1-k^2)^2$ is the inverse space
equivalent of the linear operator $r+(1+\nabla^2)^2$. The term $\hat
{NT}(\vec k,t)$ is the Fourier transform of the nonlinear part
$\psi(\vec x,t)^3+V(\vec x)$.
%In order to equilibrate we apply the
%update algorithm (\ref{eq:siupalg}) for 500 000 times with $dt=0.5$.
%%
%% KEN - more details or a reference of the algorithm should be given.
%% ENZO - ok, I leave this to Cristian or Tapio
%%
% The Laplacians are evaluated in $k$ space.
%In the presence of thermal fluctuations and the driving force, Eq.
%(\ref{coveq}) was solved using Euler's method with the Laplacians
%and gradients evaluated using finite differences. %%
%% KEN - Euler for time step?
%% ENZO - ok, I leave this to Cristian or Tapio
%% Cristian -this sentence is regarding how eq of motion was solved in the presence of the force. This was done by Enzo and Jorje.

\section{Positional and orientational correlations }

To investigate the influence of disorder on the structural
properties of the phase-field crystal, we study the behavior of the
static correlation functions from calculations of the  structure
factor $S(\vec k)$ and orientational susceptibility $\chi$, which
measure translational order and orientational order \cite{nelson},
respectively. $S(\vec k)$  can be obtained from the positions $\vec
R_j$ of the peaks in the phase-field pattern as
\begin{equation}
S(\vec k) = \frac{1}{N_p}\sum_{j,j'=1}^{N_P}  e^{-i \vec k \cdot
(\vec R_j -\vec R_{j'})} ,
\end{equation}
where $N_p$ is the number of peaks, while  $\chi$ can be obtained
from the local orientational order parameter $\phi_6(R_j)$ as
\begin{equation}
\chi = \frac{1}{N_p} \sum_{j,j'=1}^{N_p}  \phi_6(R_j)
\phi_6^*(R_{j'}).
\end{equation}
The orientational order parameter $\phi_6(R_j)$ is a measure of the
six-fold orientational symmetry of the crystalline order and is
defined as
\begin{equation}
\phi_6(R_j) = \frac{1}{N_j} \sum_{l=1}^{N_j} e^{i 6 \theta_{j,l}},
\end{equation}
where the summation is taken over the $N_j$ nearest neighbors of the
peak at position $\vec R_j$ and $\theta_{j,l}$ is the angle of the
bond $jl$ with an arbitrary axis. The disorder averaged $S(k)$  and
$\chi$ are  obtained by averaging over different realizations of the
disorder configurations $V(\vec x)$.

In the absence of disorder, solutions of the PFC model form a
periodic hexagonal array of peaks in the ground state with wavector
$\vec Q$, which has both long range positional and six-fold
orientational order. Positional order can be characterized by the
scaling of the structure factor $S(Q)$ with system size $L$, which
behaves as $S(Q) \propto L^2$ for a perfectly ordered state.
Disorder can destroy positional long range order and lead to quasi
long-range or short-range positional order, which corresponds to
correlations that decay with distance $r$ as a power law $r^{-\eta}$
or exponentially $e^{-r/\xi}$, respectively. For these distinct
types of correlations the corresponding structure factor is expected
to behave as $S(Q) \propto L^{2 -\eta}$ and $S(Q) \propto const.$,
for large system sizes. We find it convenient to define an {\it
effective exponent } $\eta_p$ to characterized the positional order
in different regimes, from a power law fit of the normalized
structure peak as
\begin{equation}
S(Q)/N_p \propto L^{-\eta_p}.
\end{equation}
With this definition, since $N_p \propto L^2$,  $\eta_p \rightarrow
0$ indicates long range order, $\eta_p \rightarrow \eta < 2 $ quasi long-range
order and $\eta_p \rightarrow 2$ short-range order. An
analogous effective exponent $\eta_o$ characterizing the
orientational correlations, can be defined from the finite-size
dependence of the orientational susceptibility $\chi$ as
\begin{equation}
\chi/N_p \propto L^{-\eta_o}.
\end{equation}

\section{Numerical Results and Discussion}
In this section, we present our numerical results for the static
behavior and the driven response of the phase-field crystal at zero
temperature (without thermal noise), obtained in the absence and
presence of the external force, respectively. The dynamical
equations, Eqs. (\ref{coveq}) and (\ref{soveq}), were integrated
numerically on a uniform square grid with $dy=dx=\pi/4$. The total
system size  $  L_x dx \times L_y dy$ accommodates approximately
$L^2/100$ density maxima.
%For the sparse
%pinning model we set the minimum distance to $L_p=a_t$, where
%$a_t=7.255$ is the lattice constant of the ideal triangular lattice
%in the absence of external potential.
The PFC parameters are set to $r=\bar \psi = -1/4$, where the ground
state of the model is a perfectly ordered hexagonal phase in the
absence of disorder, and $\gamma=1$. For the case without external
force and thermal fluctuations we used a rectangular box of size
$L_xdx\times L_ydy$ with $dy=\pi/4$ and $dx=dy/(\sqrt(3)/2)$. In
order to check the size effect we performed calculations for several
values of  $L_x$ from $128$ to $512$.

%
%%KEN - need to specify dx, dy (since they are different in
%% some simulations and dt
%%ENZO - I leave this to Cristian or Tapio. We used dx=dy=pi/4, dt=0.002
%% Cristian - should I move the part regarding the algorithm from the end of section II here?
%%

\subsection{Ground state}

We first search for the the lowest-energy state of the phase-field
crystal in the absence of the external driving force. For this
purpose, we have used the simulated annealing method to avoid
trapping in higher energy metastable states. First, simulations of
equilibrium states at some initial temperature corresponding to
finite noise terms in the dynamical equation (Eq. (\ref{coveq})) are
performed using arbitrary initial configurations. Then the initial
temperature is slowly decreased to zero leading to a final
configuration with a minimum energy. This annealing procedure is
repeated for different initial configurations and the
zero-temperature configuration with the lowest energy is identified
as the ground state.

\begin{figure}
\includegraphics[ bb= 6cm 13cm  15cm   27cm, width=7.5 cm]{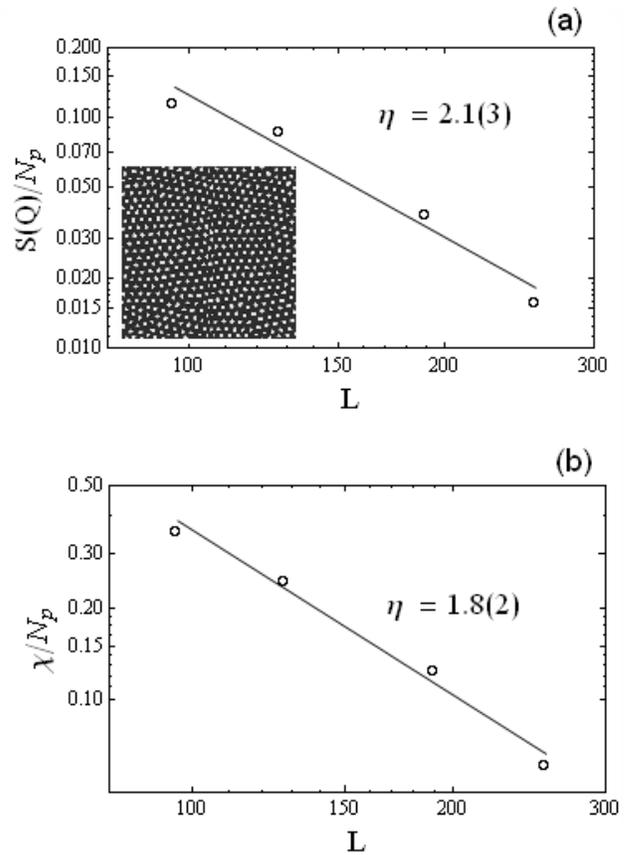}
\caption{Size dependence of the (a) structure factor peak  and (b)
orientational susceptibility, for weak disorder $D=0.03$, obtained
from simulated annealing. The straight lines are power-law fits to
the data, $S(Q)/N_p \propto L^{-\eta_p}$ and $\chi/N_p \propto
L^{-\eta_o}$ . Inset in (a): lowest-energy pattern of the phase
field $\psi(\vec x)$ for $L=190$ .}\label{sfanneal}
\end{figure}

Fig. \ref{sfanneal} shows the behavior of the structure factor peak
and orientational susceptibility as function of the system size
obtained by simulated annealing, in the weak disorder regime,
$D=0.03$.
%The inset of the Figure shows a
%phase-field pattern in the ground state for a particular disorder
%configuration. For this strength of disorder, Fig. \ref{deta}
%indicates that that the phase-field crystal would remain essentially
%ordered if no simulated annealing was performed. However, the
The effective exponent for both positional and orientational
correlations obtained from a power law fit of the size dependence in
Fig. \ref{sfanneal} is consistent with two, indicating that
correlations are short ranged. Larger disorder strength gives the
same result. This behavior suggests that the ground state of the
phase-field crystal in presence of pinning disorder is an amorphous
glass. This is expected on theoretical grounds.
%, if the system is allowed to reach
%its true ground state via simulated annealing.
%This behavior suggests that for the full equilibrated system, the
%ground state of the phase-field crystal is an amorphous glass for
%any strength of disorder with the apparent thresholds $D_{1}$
%$D_{2}$ and $D_3$ in Fig. \ref{deta} representing some
%characteristic values for non-equilibrium behavior.
In fact, analytical studies by renormalization-group methods
\cite{giamarchi95,carraro,carpentier} and computer simulations
\cite{fisher99,chandra} indicate that, in two dimensions, quasi
long-range positional and orientational order are destroyed by weak
disorder or thermal fluctuations, leading to a liquid-like phase in
the thermodynamic limit. In the absence of thermal fluctuations
(zero temperature) an amorphous-glass state is expected
\cite{carpentier,giamarchi}. The length scale for the crossover to
such state increases with decreasing disorder strength.

Long-range order in presence of weak disorder has been found at low
temperatures in some other models \cite{cha}, which however describe
structural (internal) disorder \cite{carpentier,nelson83} rather
than external pinning disorder as considered here. On the other
hand, in some molecular-dynamics simulations  of particle models of
colloidal crystals on a disordered substrate \cite{soft}, quasi
long-range order was observed at low temperatures in a weak disorder
regime, although no detailed finite-size analysis was carried out.
Quasi long-range order has also been observed in experiments on
colloidal crystals on a rough substrate \cite{Pert08}. However, the
apparent positional and orientational order in such cases can also
be explained as a finite-size effect due to a large but finite
crossover length scale, which increases with decreasing disorder
strength \cite{carpentier,doussal00}. Another possible explanation
%is that it may simply take too long to reach an equilibrium state.
is the effect of slow dynamics, which requires a much longer time
scale to observe the true equilibrium state, leading to
partially-ordered metastable states. In Sec. IVc, we examine similar
states for the PFC model, which were obtained using the dynamical
equations in absence of thermal annealing.

\subsection{Behavior under a driving force}

\begin{figure}
\includegraphics[ bb= 6cm 20cm  15cm   28cm, width=7.5 cm]{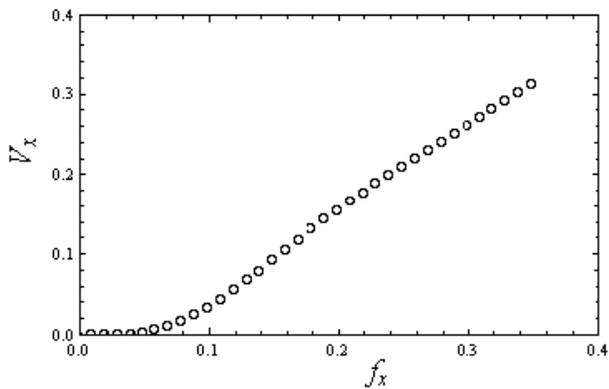}
\caption{Longitudinal velocity response $V_x$ to a driving force
$f_x$ averaged over disorder, for $D=0.1$ and $L=128$. }\label{vxf}
\end{figure}

\begin{figure}
\includegraphics[ bb= 3cm 9cm  18cm   28cm, width=7.5 cm]{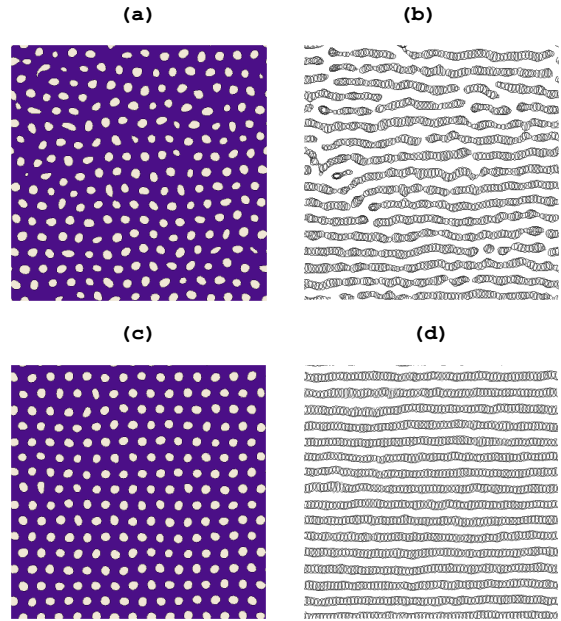}
\caption{(Color online) Snapshot of the phase field $\psi(\vec x)$
and corresponding trajectories of the peaks in $\psi(\vec x)$ in the
moving state. (a) and (b) for $f=0.2$  and (c) and (d) for $f=0.8$.
Results are  for $L=128$ and $D=0.1$. The trajectories in (b) and
(d)) are superpositions of snapshots of the peaks (open circles in
(a) and (c)) at consecutive times. }\label{patf02f08}
\end{figure}

\begin{figure}
\includegraphics[ bb= 5cm 11cm  16cm   28cm, width=7.5 cm]{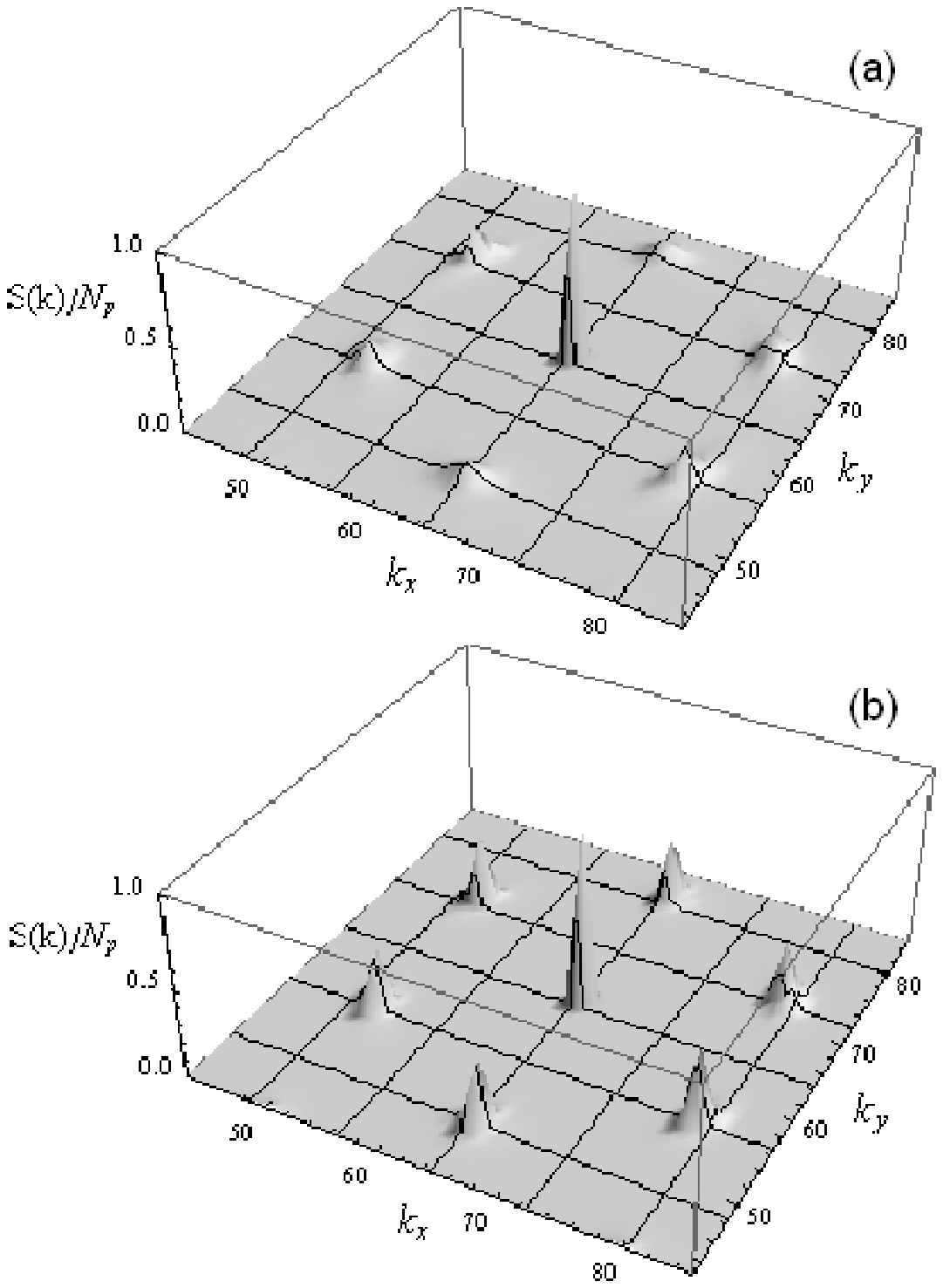}
\caption{Structure factor, averaged of the disorder, in the moving
state, for (a)  $f=0.2$  and (b) $f=0.8$. Results are for $L=128$
and $D=0.1$. }\label{sff02f08}
\end{figure}

\begin{figure}
\includegraphics[ bb= 5cm 17cm  16cm   26cm, width=7.5 cm]{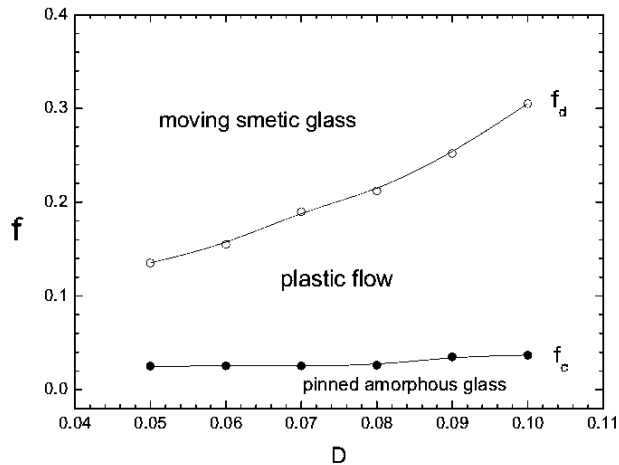}
\caption{Qualitative dynamical phase diagram as a function of
disorder strength $D$ and applied force $f$. $f_c$ is the depinning
force and $f_d$ is the critical value for the transition from
plastic flow to moving glass regimes.}\label{phadiag}
\end{figure}

To obtain the driven response of solutions of the phase-field
crystal model, we determine the steady state velocity of the peaks
under a constant external force as described by Eq. (\ref{coveq}).
%(KEN 5 or 6?).
The velocity is measured by
tracking the position
of the peaks $\vec R_i(t)$ in the phase field $\psi(\vec x)$ during
the simulation \cite{achim09}. From the velocities of the peak
positions $\vec v_i=d\vec R_i/dt$, the steady state drift velocity
is obtained  as
\begin{equation}
\vec v = \langle \frac{1}{N_P}\sum_{i=1}^{N_P} \vec v_i(t) \rangle,
\label{velpeak}
\end{equation}
where $N_P$ is the number of peaks and $\langle ... \rangle$ denotes
time average.  The disorder averaged drift velocity is then obtained
by averaging  over different realizations of the disorder
configurations. The calculations were done without thermal noise
(zero temperature). The initial state at $f=0$ is obtained from
simulated annealing as described in Sec. IVa. The velocity response
of the phase-field crystal to the applied driving force is shown
Fig. \ref{vxf} for strong disorder $D=0.1$. The velocity response is
nonlinear, with different behavior at low and large driving forces,
due to the effects of pinning. This property is observed in
different lattice systems with quenched disorder
\cite{giamarchi,balents,koshelev}. At sufficiently low drive, the
drift velocity is negligibly small and the phase-field crystal
remains pinned in the amorphous glassy phase. As the force increases
beyond a critical value, $f_c \approx 0.05$, %%
%% KEN: from the figure it looks like depinning occurs at a higher
%% force - closer to 0.2 ?? Does it really move at .1?
%% ENZO - in a enlarged velocity scale, 0.1 is a better estimate
%% move here means V>0. It is difficult to see the peaks moving when V~0.
%%
corresponding to a depinning transition, the phase-field crystal
moves with increasing steady-state velocity but with different
patterns at low and high velocities as can be seen from the
corresponding configurations in Fig. \ref{patf02f08}. At low
velocities above the critical force $f_c$, the pattern of the phase
field  has a liquid-like structure as shown in Fig. \ref{patf02f08}a
for $f=0.2$. The trajectories of the local maxima in the phase field
$\psi(\vec x)$ correspond to a plastic flow due to pinned and
unpinned regions. For this moving state, the corresponding structure
factor (Fig. \ref{sff02f08}), averaged over time and disorder, shows
small and broad peaks. For larger values of the driving force above
a characteristic value $f_d$, there is a dynamical reordering
transition to a moving partially ordered phase as shown in Fig.
\ref{patf02f08}c for $f=0.8$. The trajectories of the local maxima
form well defined static channels parallel to the driving force. The
structure factor now become larger and sharper as shown in Fig.
\ref{sff02f08}. Rough estimates of critical values $f_c$ and $f_d$
obtained for different disorder strengths were used to construct the
qualitative dynamical phase diagram as a function of disorder
strength and applied force shown in Fig. \ref{phadiag}.

\begin{figure}
\includegraphics[ bb= 6cm 13cm  16cm   28cm, width=7.5 cm]{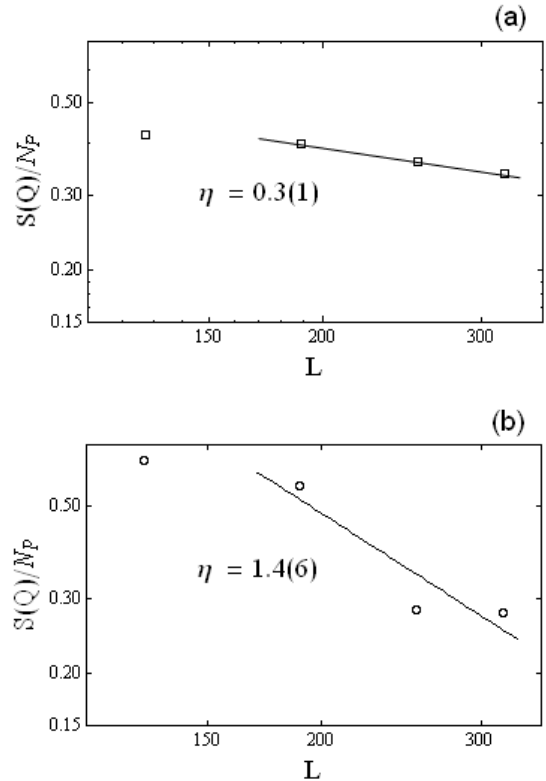}
\caption{Finite-size behavior of the (a) transverse and (b)
longitudinal peaks in the structure factor, for the moving state at
$f_x=0.8$. The straight lines are power-law fits to the data,
$S(Q)/N_p\propto L^{-\eta}$, for the largest sizes. }\label{sff08}
\end{figure}

The nature of the moving phase at high velocities is particularly
interesting.  Since the pinning potential acts as a time oscillating
perturbation in the co-moving reference frame, one would expects
that the disorder effects on the initial hexagonal structure should
decrease with increasing velocity. However, it has been shown that
some components of the disorder remain static \cite{giamarchi},
leading to a moving glass phase. In two dimensions, such a phase
corresponds to a moving transverse glass characterized by a
smectic-like  structure, which retains quasi long-range order in the
direction perpendicular to the applied force but only exponential
correlations in the direction parallel to the force
\cite{giamarchi,balents}. This moving smectic state has been
observed in many different driven systems with disorder
\cite{Moon,olson98,olson00,fanghor,reichardt,lefebfre}. Indeed, for
the present driven PFC model, the pattern for the large driving
force in Fig. \ref{patf02f08}d also shows a smectic-like structure.
To verify this behavior more quantitatively, we have studied the
finite-size dependence of the structure factor peaks (Fig.
\ref{sff02f08}) in the transverse and longitudinal directions for
the sliding state at $f=0.8$. The results shown in a log-log plot in
Fig. \ref{sff08} are consistent with a smectic phase in the
thermodynamic limit, where ordering only occurs in one spatial
direction. The height of the peaks in the transverse direction
decreases slowly with system size as a power law $ S/N_P^2 \propto
L^{-\eta}$ with $\eta \approx 0.3(1)$, consistent with quasi
long-range order, while in the longitudinal direction it decreases
much faster with $\eta \approx 1.4(6)$ as expected for short-range
order. This estimate of the critical exponent $\eta $ for the
transverse peaks in the driven phase-field crystal is comparable to
the result, $\eta =0.5(1)$, obtained for a particle model of driven
vortex lattices in disordered superconductors \cite{Moon}.

\begin{figure}
\includegraphics[ bb= 4cm 17cm  17.5cm   27cm, width=7.5 cm]{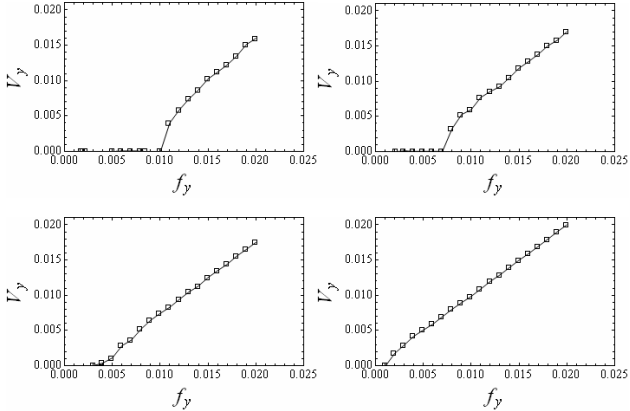}
\caption{Transverse velocity response $V_y$ to an additional force
$f_y$ for different disorder configurations, for the moving state at
$f_x=0.8$ when $D=0.1$ and $L=128$. }\label{trans}
\end{figure}

\begin{figure}
\includegraphics[ bb= 6cm 20.5cm  16cm   27cm, width=7.5 cm]{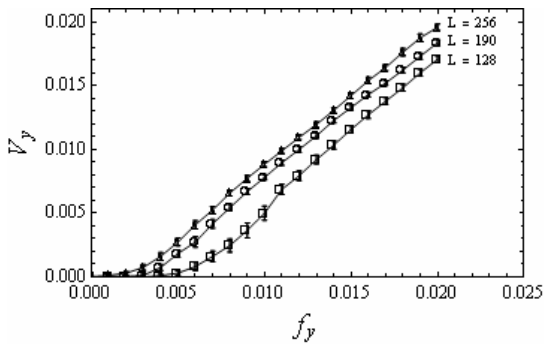}
\caption{Finite-size dependence of the transverse velocity response
$V_y$ to an additional force $f_y$ averaged over disorder, for the
moving state at $f_x=0.8$ and $D=0.1$. }\label{avtrans}
\end{figure}

Another important property of the moving glass state at large
velocities is the existence of barriers to transverse motion. It was
predicted \cite{giamarchi} that this should lead to a transverse
critical force at zero temperature for an additional external force
applied perpendicular to the initially driving force. However, it
has also been argued \cite {balents}  that the transverse response
at zero temperature is nonuniversal, and  qualitatively different
behavior is possible for different physical systems. The former
scenario of transverse depinning has already been observed for
driven vortex lattices
\cite{Moon,olson98,olson00,fanghor,reichardt,lefebfre}. To find out
which scenario is realized in the present version of the PFC model,
we have studied the velocity response in the transverse direction
$V_y$ at a large longitudinal force $f_x$ for different realizations
of the disorder configurations. We find that the transverse
depinning force is very sensitive to the disorder configurations and
seems to vanish in some cases as can be seen in Fig. \ref{trans},
where the transverse velocity response is shown for different
realizations of the disorder configuration for a system size
$L=128$. To investigate if the transverse critical force is still
nonzero in the large system limit, we performed calculations for
increasing system sizes and the results were averaged over different
disorder configurations.  Fig. \ref{avtrans} shows that the critical
value $f_{cy}$ for the onset of transverse depinning decreases with
increasing system size and therefore may actually vanish in the
thermodynamic limit. Although such behavior is possible according to
analytical arguments \cite{balents}, it is unclear at the present
which distinct feature of the PFC model is responsible for it. Since
a transverse critical force has been found for the same model in the
case of periodic pinning \cite{achim09}, we can only speculate that
the presence of disorder allows some defects to be generated even by
small transverse forces.

\subsection{Glassy metastable states}

In the previous  section IVa  we have determined the lowest energy
state in absence of a driving force (ground state) using simulated
annealing methods and found that the configuration corresponds to an
amorphous glass state, with only short-ranged positional and
orientational correlations.
%The response of this state to an external driving force was then
%studied.
In this section, we will discuss some metastable states. These are
states corresponding to local minima rather than the global minimum
for the ground state. Nonetheless, they may still be relevant under
suitable experimental conditions.

Metastable states were obtained by starting with the perfect
hexagonal ground state in the absence of disorder and then allowing
the system to evolve according to the dynamical equation for a
relatively short period ($5\times 10^5 dt, dt=0.5 $ ), for different
disorder strengths in the  absence of thermal noise. The disorder
strength $D$ is incremented in steps of $0.009$.

\begin{figure}
\includegraphics[ bb= 4cm 10cm  17cm   27cm, width=7.5 cm]{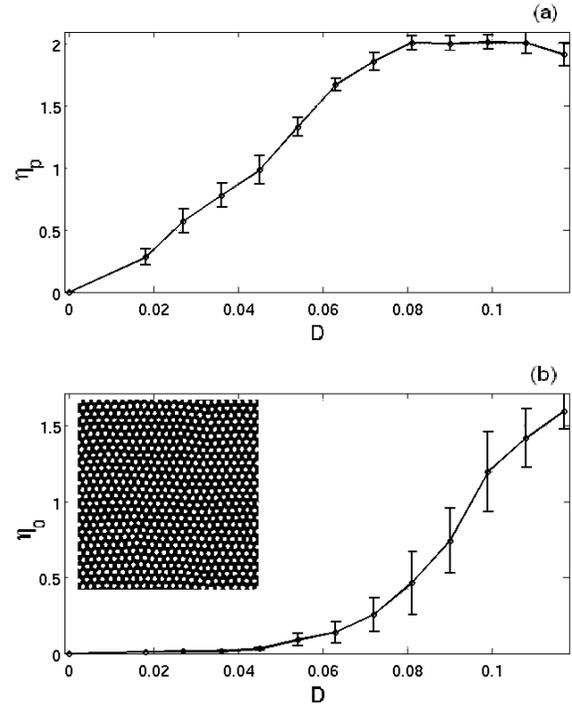}
\caption{Effective power-law exponents for (a) positional $\eta_p$
and (b) orientational $\eta_o$ correlation functions, obtained from
the dynamical equations without thermal noise. Inset in (b): phase
field pattern for $D=0.036$. }\label{deta}
\end{figure}

\begin{figure}
\includegraphics[ bb= 4cm 10cm  17cm   27cm, width=7.5 cm]{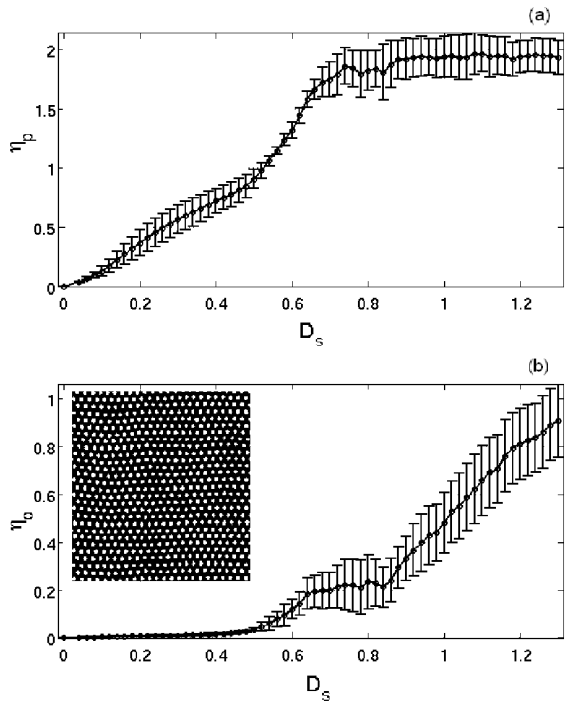}
\caption{Effective power-law exponents for (a) positional $\eta_p$
and (b) orientational $\eta_o$ correlation functions in the {\it
sparse} pinning model (see text for details), obtained from the
dynamical equations without thermal noise. Inset in (b): phase field
pattern for $D_s=0.4$.}\label{seta}
\end{figure}

Fig. \ref{deta} shows the behavior of the effective exponents
$\eta_p$ and $\eta_o$ for positional and orientational correlations
of the final static configuration as a function of the disorder
strength $D$. It appears that there are three different thresholds
values of $D$: $D_1 \approx 0.04$, $D_2 \approx 0.08$ and $D_3 >>
0.12$.  Below $D_1$, the {\it weak-disorder} regime,  the system has
long-range orientational order and quasi-long range translational
order. Between $D_1$ and $D_2$ the system exhibits both quasi-long
range orientational and translation order. Between $D_2$ and $D_3$ a
hexatic ordering occurs which corresponds to quasi-long range
orientational order and short-range translational order. This phase
ordering is analogous to such a phase that occurs in two dimensional
crystals induced by thermal flucuations \cite{nelson}. Finally above
$D_3$, which is beyond the range of $D$ investigated, we expect that
only short-range orientational and translational order should
remain.  As described in Sec. IVa, the lowest-energy state obtained
by simulated annealing is an amorphous glass rather than a
partially-ordered state observed in such netastable states. In fact,
the Inset in Fig. \ref{deta} indicates that the phase-field crystal
would remain essentially ordered for weak pinning if no simulated
annealing was performed. However, even in this weak disorder regime
correlations are actually short ranged if the system is allowed to
reach its true ground state, as can be seen from the Inset in Fig.
\ref{sfanneal}. Although the ground state is an amorphous glass, the
thresholds $D_{1}$ $D_{2}$ and $D_3$ in Fig. \ref{deta} can
represent some characteristic values for non-equilibrium behavior.

We have also investigated the metastable states for a different
model of disorder. In this {\it sparse} pinning model, the pinning
sites are separated by a minimum distance $L_p \approx 2\pi/(|\vec
Q|\sqrt(3)/2)) $, {\it i.e.} they cannot come closer to each other
than the particles in the ideal hexagonal ground state of the PFC
model with wave vector $\vec Q$. Such model may be realized {\it
e.g.} in the case of adsorbed atomic layers on a substrate with
impurities \cite{Book}. A finite density $\rho_s$ of randomly
distributed pinning centers is assigned with pinning strength
(amplitude) $D_s$. Thus, unlike the {\it dense} pinning model
discussed above, there are now two relevant parameters, namely
$\rho_s$ and $D_s$. Metastable states were obtained again starting
from the hexagonal state and incrementing $D_s$ in steps of $0.02$,
while the density of pinning sites $\rho_s$ was fixed to
$\rho_s=0.0082$. For this density of the pinning centers the ratio
between the pinning centers and the number of maxima of the ideal
triangular lattice is $0.375$. Fig. \ref{seta} shows the behavior of
the effective exponents $\eta_p$ and $\eta_o$ for positional and
orientational correlations of the final static configuration as a
function of the disorder strength $D_s$ obtained in this sparse
disorder model. Although the errorbars are relatively large, there
is again three distinct thresholds values, $ D_{1}\approx 0.45 $,
$D_2 \approx 0.7$ and presumably $D_{3} >> 1.3$. The results are
qualitatively the same as that for the dense pinning model.

\section{Conclusions}

We have studied  a two-dimensional PFC model with random pinning.
The model provides an effective continuous description of lattice
systems in the presence of  disordered external pinning centers,
allowing for both elastic and plastic deformations. The structural
correlations and nonlinear driven response are determined from
numerical simulations of the dynamical equations. We find that in
the absence of a driving force, the phase-field crystal assumes an
amorphous glassy ground state, with short-ranged positional and
orientational correlations even in the limit of weak disorder.
%However, initial ordered states can evolve into partially-ordered
%states in presence of pinning disorder, in absence of thermal
%annealing. are found when
Under increasing driving force, this pinned amorphous-glass phase
evolves first into a moving plastic-flow phase then followed by a
moving smectic phase. These results are largely in agreement with
previous analytical works \cite{giamarchi95,carraro,carpentier}. An
additional new feature is that the transverse response of the moving
smectic phase shows a vanishing transverse critical force for
increasing system sizes. Finally, we have identified interesting
quasi-long range ordered  metastable states by evolving initially
ordered states in the presence of disorder. The nature of these
metastable states are insensitive to the details of disorder and
they may be relevant under suitable experimental conditions.

\begin{acknowledgments}

E.G. was supported by Funda\c c\~ao de Amparo \`a Pesquisa do Estado
de S\~ao Paulo - FAPESP (Grant No. 07/08492-9). K.R.E. acknowledges
support from the NSF under Grant No. DMR-0906676. This work was also
supported in part by computer facilities from the Centro Nacional de
Processamento de Alto Desempenho ( CENAPAD-SP and CENAPAD-UFPE ).

\end{acknowledgments}

\end{document}